\documentclass[fdp,a4paper,fleqn%
]{w-art}
\usepackage{times,cite,w-thm}
\usepackage{amsmath}
\usepackage{amssymb}
\usepackage{mathrsfs}
\newcommand{\IC}{\ensuremath{\mathbb {C}}}
\newcommand{\IP}{\ensuremath{\mathbb {P}}}
\newcommand{\IZ}{\ensuremath{\mathbb {Z}}}
\newcommand{\Tr}{{\rm Tr}}
\newcommand{\beq}{\begin{equation}}
\newcommand{\eeq}{\end{equation}}
\newcommand{\bea}{\begin{eqnarray}}
\newcommand{\eea}{\end{eqnarray}}
\theoremstyle{plain}

\theoremstyle{definition}

\usepackage[]{graphicx}
\begin{document}
\DOIsuffix{theDOIsuffix}
\Volume{55}
\Month{01}
\Year{2007}
\pagespan{1}{}





\title[Membranes and Integrability]{Membranes, Strings and Integrability}


\author[C. Krishnan]{Chethan Krishnan\inst{1,}%
  \footnote{
E-mail:~\textsf{chethan.krishnan@ulb.ac.be}
            }}
\address[\inst{1}]{International Solvay Institutes, Physique Th\'eorique et Math\'ematique, \\
ULB C.P. 231, Universit\'e Libre de Bruxelles, B-1050, Bruxelles, Belgium}
\author[C. Maccaferri]{Carlo
Maccaferri\inst{1,}\footnote{E-mail:~\textsf{cmaccafe@ulb.ac.be}}}
\begin{abstract}
In the first half of this note, after briefly motivating and reviewing membrane field theories, we consider their BPS funnel solutions. We discuss some aspects of  embedding M-theory fuzzy funnels in these theories. In the second half, we focus on ABJM theory and explain a test of $AdS_4/CFT_3$ based on integrability. We discuss a numerical mismatch at one loop in worldsheet perturbation theory and its possible resolutions.
\end{abstract}
\maketitle                   





\section{Introduction}

Strongly coupled type IIA string theory at low energies reduces to eleven dimensional supergravity. This is part of the motivation for M-theory, the proposed UV completion of 11D SUGRA. The trouble is that such an indirect definition is often not good enough to give us full control on the questions that we would like to answer, in particular, regarding the vacuum structure of the theory. What we would like instead, is to have a microscopic description of M-theory. Some attempts at a frontal attack of this problem have been made (starting from the ground-breaking work of \cite{BFSS}), but these methods are usually tied to flat spacetime and its relatives. Having other inroads into the microscopics of M-theory would certainly be a major advance.

In the case of string theory, D-branes (which are non-perturbative states in the theory) offered us a window into many strong coupling effects. This was essentially because D-branes could be studied in two separate ways. They could be thought of either as classical solutions of low energy closed string theory or as effective descriptions of the endpoints of open strings. This duality is at the core of, for example, the AdS/CFT correspondence which gave us tremendous insights into both gauge theory and string theory. Unfortunately, the lack of a microscopic description for M-theory has prevented us from explicitly constructing the worldvolume theories on M-branes, in contradistinction to the case of D-branes where worldsheet open string theory is a useful tool.

As often in physics, one can get far by using symmetry arguments alone, even if the detailed dynamics is not under control. One thing we do know about membranes \cite{DB-review} is that in the closed string language, they should correspond (in the large radius regime) to membrane solutions of 11D SUGRA. The near-horizon region of such solutions gives rise to an $AdS_4 \times S^7$ flux background. Coupled with the knowledge that membranes are $2+1$ dimensional and that they break half of the 32 supersymmetries, the symmetries of $AdS_4 \times S^7$ tell us that the low energy worldvolume theory on multiple M2-branes should be a $2+1$ dimensional ${\cal N}=8$ superconformal theory with an $SO(8)$ R-symmetry.

Is it possible to construct a Lagrangian description of membrane field theories that manifests these features? Despite the general belief otherwise, Bagger, Lambert \cite{Bagger} and Gustavsson \cite{Gustavsson} showed that the answer is in fact ``yes". They constructed an action for multiple membranes using the so-called 3-algebras. This theory 
had all the properties expected from membrane worldvolume theories bar one: there is no free parameter in the theory that can be interpreted as the number of M2-branes. After a lot of intense work that lead to many directions of progress, the concensus now is that BLG theory is not a general theory of any number of membranes, but that it is closely related to a specific configuration of two membranes probing an orbifold \cite{Tong, ABJM}. In the first half of this article, we will investigate some of the BPS states of BLG theory and see that it has enough structure to incorporate the most general fuzzy funnels of M-theory, where M2-branes expand into intersecting configurations of M5-branes \cite{KM}. Of course, since the number of branes in BLG theory is only two, the number counting of branes is not satisfactory.  Efforts to generalize BLG theory to include more branes within the original framework of 3-algebras was attempted in \cite{ABC}, but we will show that these theories are not promising candidates for producing fuzzy funnels. There have also been other arguments in the literature which suggest that these ``negative trace form" theories are nothing more than a rewriting of ${\cal N}=8$ super Yang-Mills theory in 2+1 dimensions \cite{Mukhi}, whereas what we are really after is an explicit description of the IR fixed point of ${\cal N}=8$ SYM. So in the second half of this article, where our intention is to test $AdS_4/CFT_3$, we will turn our attention instead to ABJM theory \cite{ABJM}, a different generalization of BLG theory without directly resorting to 3-algebras\footnote{But see \cite{bl} for a connection of ABJM theory to 3-algebras.}. We will adopt the viewpoint that ABJM is the most promising candidate available in the market currently for a theory of many membranes.

Since ABJM theory is supposed to be a generic theory of membranes, it involves both a coupling (the Chern-Simons level) and a large $N$ limit (rank of gauge groups). By tuning these parameters, we can relate it to a type IIA string theory on $AdS_4 \times \IC \IP^3$ where explicit gauge-string comparisons are possible. In particular, it has been shown that ABJM is integrable \cite{Minahan}\footnote{The worldsheet string theory was also shown to be classically integrable in \cite{stefanski}. The conformality of the background to all loops was shown using pure spinors in \cite{purespinor}. Giant magnons in $AdS_4/CFT_3$ have been considered in \cite{Orselli}.}, and a BES-type \cite{BES} conjecture was made on the gauge theory side \cite{GV}. This enables a strong coupling expansion to be performed on the gauge theory side which should be comparable against worldsheet perturbation theory. We will find that even though the gauge-string results match in form, numerical values are different at one loop in the sigma model\cite{MR, Fernando, CK}. This result shows that there are some subtleties in the gauge-string comparison in $AdS_4/CFT_3$ which were not present in the more familiar case of integrable $AdS_5/CFT_4$. We will conclude with some comments about possible resolutions of this discrepancy on the gauge theory Bethe ansatz side \cite{MRT}. Our focus in this short note will be to report results: the reader should consult the references cited for more details and background material.

\section{A ``Derivation" of Bagger-Lambert Theory}

We are interested in certain BPS solutions of Bagger-Lambert theory which can be interpreted as fuzzy funnels of M-theory, known from the work of \cite{Shahin, Basu, Berman}. To work up to it, we will first start with general arguments about the supersymmetries preserved by stacks of M2-branes, since these are ultimately what lead to the BPS equations. 

In the case of D-branes, the worldvolume theory is described by the transverse scalars, $X^I$ where $X^I$ are elevated to {\em matrices}. So for the case of M2's, we can start by trying to write a theory for $X^I_a$ where $I=(3,...,10)$ are the transverse directions and $a$ is a (multi-)index. 
From balancing various indices on either side, one can see that the most general (linear) way in which the 16 unbroken SUSY's can act is as
\bea
\delta X^I_a=i \bar \epsilon \ \Gamma^I  \ \Psi_a, \ \  {\rm with} \  \ \epsilon=\Gamma_{012}\epsilon.
\eea
If we assume canonical kinetic terms for the spinors and the scalars, in $2+1$ dimensions, we have
$[X]=\frac{1}{2} \  {\rm and} \  [\Psi_a]=1$,
as the scaling dimensions of the fields. With a bit of trial and error, it is easy to convince oneself that this means that the most general (without adding extra fields) SUSY variation that one can write down consistent with balancing spinor indices, internal indices and dimensions on either side is
\bea
\delta \Psi_a=\partial_\mu X^I_a \Gamma^{\mu I} \epsilon + c \ X^I_b X^J_c X^K_d f^{bcd}_{ \ \ \ \ a}\Gamma^{IJK} \epsilon
\eea
where $c$ is a parameter and the $f^{bcd}_{ \ \ \ \ a}$ are ``structure constants". The crucial observation of Bagger, Lambert and Gustavsson was to note that to close such a SUSY variation, one needs to covariantize the derivative  $\partial_\mu X^I_a$ in the above expression by introducing a gauge field. The rest follows more or less automatically upon demanding closure of SUSY on shell: {\bf (1)} The parameter $c$ gets fixed to $-\frac{1}{6}$, {\bf (2)} The structure constants $f^{bcd}_{ \ \ \ \ a}$ are to satisfy the so-called {\em fundamental identity}: $f^{[abc}_{ \ \ \ \ g}f^{e]fg}_{ \ \ \ \ d}=0$,
{\bf (3)} The equations of motion of the various fields are fixed.

The EOMs arising from the closure of the algebra can be obtained from an action. This is the BLG action (we will not write it down in full glory). But to construct that action, we need to assume two crucial things: {\bf (1)} The existence of a trace form $h^{ab}$ which can be used to raise indices so that we can construct scalars, {\bf (2)} $f^{abcd}\equiv h^{de}f^{abc}_{ \ \ \ \ e}$ is fully antisymmetric in all indices.
Unfortunately, if one restricts to positive definite $h^{ab}$, the only solutions to these restrictions is given by $f^{abcd}=\epsilon^{abcd}$ \cite{Bandres}. This choice is what corresponds to the original BLG theory. We will say some words about indefinite $h^{ab}$ in the next section.

\section{BPS Funnels}

We will start with the example of a fuzzy funnel from string theory : the BIon \cite{bion}. These are solutions of the BPS equations in the worldvolume theory of D1-branes, which expand into D3-branes. In general these configurations can expand into intersecting configurations of D3 branes \cite{const}. We wish to see the emergence of such configurations in the case of M2s expanding into M5s. They were constructed through inspired guesswork before the emergence of Bagger-Lambert theory in \cite{Shahin, Basu} and generalized by Berman-Copland \cite{Berman}. (See also \cite{Bonelli} for more recent work. Non-linear memebrane actions which might be able to reproduce fuzzy funnels have been considered in \cite{non-linear}.)

To see funnels in Bagger-Lambert theory, we will write the scalar part of the BLG action
\bea
\label{action}
{\cal L}_B=-\frac{1}{2}\Tr\,(\partial_\mu X^I,\partial^\mu X^I)
-\frac{1}{12}{\rm Tr}([X^I,X^J,X^K],[X^I,X^J,X^K])
\eea
where $X^I \equiv X^{I}_a\,T^a$, $[T^a,T^b,T^c] = f^{abc}{}_{d}\,T^d$, and $h^{ab}=\Tr\,(T^a,T^b)$ for the ``3-algebra" generators $T^a$. The BPS funnels arise when we set the energy functional computed from this Lagrangian to zero. This is because $Q |\psi\rangle =0$ implies $\langle \psi| \{Q, Q\} |\psi \rangle \sim \langle\psi| H|\psi \rangle =0$. Splitting off a total derivative piece from the Hamiltonian, we can write (for appropriate coefficients $g_{IJKL}$)
\bea
E=\frac{1}{2}\int d^2 \sigma \left( {\rm Tr}\left(\partial_\sigma
X^I-\frac{g_{IJKL}}{3!}
[X^J,X^K,X^L]\right)^2+T \right).
\eea
It is possible to write the Hamiltonian this way, if the configuration satisfies certain algebraic constraints in terms of the $X$'s\footnote{These constraints can also be viewed as arising from the consistency between the BPS equation and the equation of motion.}. For the calibrated intersections of M5-branes that the M2s can expand into, these constraints are automatically satisfied in BLG theory \cite{KM} due to the fundamental identity. It can also be checked that the other equations of motion arising in Bagger-Lambert, which where not visible in the {\em ad-hoc} constructions, are also satisfied\cite{KM}. Therefore we can consistently read off the first piece in the expression above as the fuzzy funnel equations of Basu-Harvey and Berman-Copland. Solutions of the BPS equation can be found by solving the auxiliary algebraic equation 
$\frac{1}{6}g_{IJKL}[A^J,A^K,A^L]=A^I$,
because then $X^I(\sigma)=f(\sigma)A^I$ is a solution for $f(\sigma)$ satisfying $\partial_\sigma f(\sigma) = f^3(\sigma)$. The fuzzy funnels found in the literature can be constructed by a suitable definition of the 3-algebra using fuzzy 3-spheres \cite{Berman, KM}. As mentioned in the introduction, however, in BLG theory the number of membranes is only two, so we need a more general theory of many membranes to have a complete picture of M-theory funnels.

One crucial ingredient in our Bogomolnyi positivity argument above is that it works only if the trace form $h^{ab}$ is positive definite, because otherwise the energy is unbounded below. This means that attempts to generalize BLG theory by relaxing this positivity  do not fit into this picture. In negative trace form theories, the energy functional instead takes the form \cite{ABC},
\bea
\mathcal{H} = \frac{1}{2}{\rm Tr}\Big(\partial_{\sigma}X^I \partial_{\sigma}X^I\Big) - \partial_{\sigma}X^I_+\partial_{\sigma}X_-^I + \frac{1}{12} {\rm Tr}\Big(X_+^I [ X^J, X^K] + ... \Big)^2
\eea
The dots represent cycling the $I, J, K$ indices. This expression is written after expanding the 3-algebra expressions in terms of ordinary Lie algebras (from which the 3-algebras are constructed in these theories). In particular, the trace above is the usual trace of the Lie algebra and therefore positive definite. The negative trace of the 3-algebra gives rise to the negative sign of the $\partial_{\sigma}X^I_+\partial_{\sigma}X_-^I$ term. The $X_-^I$ is a Lagrange multiplier term enforcing the condition $\partial_\sigma^2 X^I_+=0$. If we solve for it by $X^I_+ \sim \sqrt{\lambda}$ with $\lambda > 0$, then the Hamiltonian is schematically that of a $\lambda \phi^4$ theory, and is positive definite. But the structure now looks like
$\ {\cal H} \sim (\partial_\sigma X + [X,X])^2\ $, which is suggestive of D2-D4 fuzzy 2-funnel intersections in Yang-Mills theory, whereas we need something like
$\ {\cal H} \sim (\partial_\sigma X + [X,X,X])^2\ $
to get fuzzy 3-funnels that connect M2s to M5s. In particular, we need three extra dimensions. It is perhaps possible that the theory can contain fuzzy 3-funnels which are realized in some non-linear way, but we will not pursue this possibility here and switch gears in the next section to ABJM theory, and to testing $AdS_4/CFT_3$ using its integrability.

\section{Chern-Simons Theories for Membranes and their Gravity Duals}

An approach that has been fruitful in constructing multiple membrane theories is to look at Bagger-Lambert theory not in terms of 3-algebras, but as a Chern-Simons theory with two gauge groups, coupled to bifundamental matter \cite{Raam}. This was done by ABJM \cite{ABJM}, and they proposed that $N$ M2-branes probing the orbifold $\IC^4 /\IZ_k$ is described at low energies by a $U(N)_k \times U(N)_{-k}$ Chern-Simons theory where $(k,-k)$ are the Chern-Simons levels of the two gauge groups. The theory has only ${\cal N}=6$ manifest supersymmetry, but is superconformal, and has bifundamental matter and a specific quartic superpotential. The inverse Chern-Simons level $1/k$ acts as the coupling of the theory, and $N/k$ is the 't hooft coupling.
The membranes are in flat space when $k=1$ but then the theory is hopelessly strongly coupled. ABJM is related to BLG when $N=2$, but the moduli spaces are not quite the same because there are extra $U(1)$ factors in the theory.

The gravity dual of ABJM is given by the near horizon limit of flat space times the orbifold, which gives us M-theory on $AdS_4 \times S^7/\IZ_k$. The $S^7$ is a Hopf fibration of a circle on $\IC \IP^3$ (see e.g., \cite{example} for circle fibrations of this kind), $\IC \IP^3$ has an $SU(4)$ isometry, and the orbifold action $z_i \rightarrow e^{\frac{2\pi i}{k}}z_i$ preserves $SU(4) \times U(1)$, so this gives rise to
\bea
ds^2_{S^7/\IZ_k}=ds^2_{\IC \IP^3}+\Big(\frac{d\phi}{k}+\omega\Big)^2.
\eea
In the large $Nk$ limit, the $\IC \IP^3$ of the near-horizon metric
\bea
ds^2=\frac{R^2}{4}ds^2_{AdS_4}+R^2ds^2_{S^7/\IZ_k}
\eea
remains large because $R\sim (Nk)^{\frac{1}{6}}$ (see the explicit 11D sugra solution \cite{ABJM}). But because of the fibration written above, it is clear that we can tune $k$ to make the circle small simultaneously. Treating this as the M-theory circle, we find that we have a type IIA description on $AdS_4\times \IC \IP^3$.

\section{Integrability and a Test of $AdS_4/CFT_3$}

An interesting feature of ABJM gauge theory is that it is integrable in the scalar sector\cite{Minahan}, which means for our purposes here, that it is possible to compute the anomalous dimensions of certain operators even at strong coupling\cite{GV}. An inverse-coupling expansion at strong gauge coupling should be reproducible by a worldsheet computation on the string side by the AdS/CFT correspondence. So this gives us a window to test $AdS_4/CFT_3$, or if one believes AdS/CFT, to sharpen the ansatzes that are used to integrate (i.e., solve) the gauge theory.

By strong coupling above, we mean strong 't Hooft coupling at large $N$. This means that only planar contributions arise, and that we are looking at worldsheet perturbation theory on a sphere. Following the example of $AdS_5/CFT_4$ \cite{Klebanov, Tseytlin}, we will consider a class of string states which have large angular momentum in $AdS_4$ (``Frolov-Tseytlin Sector")\footnote{classical solutions of strings in $AdS_4\times \IC \IP^3$ have been considered in e.g., \cite{Lee}.}. The dual operators of these states in ABJM theory can be identified (see \cite{alday2} for a discussion of their scaling in various theories), and their anomalous dimensions have been computed using the integrability of the gauge theory \cite{GV}:
\bea
\label{bes}
E-S=f_{{\rm CS}}(\lambda) \ \ln S, \ \ \ \  {\rm where} \ \ \ \ f_{{\rm CS}}(\lambda)=\sqrt{2\lambda}-\frac{3 \ln 2}{2 \pi} + {\cal O}\Big(\frac{1}{\sqrt{\lambda}}\Big)
\eea
The function $f(\lambda)$ is called the cusp-anomaly.
In the dual gauge theory, the AdS energy translates to the dimension of the operator and the angular momentum translates to the spin and this is the motivation for the choice of notation. Our aim is to reproduce this on the string theory side with a one loop string computation. To avoid suspense: the result \cite{CK} agrees in form with the gauge theory, but the precise value of the cusp-anomalous dimension is different at one loop:
\bea
\label{string}
f_{{\rm string}}(\lambda)=\sqrt{2\lambda}-\frac{5 \ln 2}{2 \pi} + {\cal O}\Big(\frac{1}{\sqrt{\lambda}}\Big)
\eea
In the following we will sketch some aspects of this result.

Computing this function on the string side is essentially about computing the energy of the spinning string at one loop in sigma model perturbation theory. We look at IIA sigma model with target\footnote{The $R$ here is different from the $R$ in the 11D theory of the last section. Also, we set $\alpha'=1$.}
\bea
ds_{IIA}^2=R^2(ds_{AdS4}^2+4ds_{\IC\IP3}^2), 
\eea
and the classical spinning string corresponds to taking $t=\kappa \tau, \phi=\omega \tau, \rho (\sigma)=\rho(\sigma+2\pi)$ where $(\tau,\sigma)$ are worldsheet coordinates and $t, \rho$ and $\phi$ correspond to an $AdS_3$ subspace of $AdS_4$. The equations of motion, the Virasoro constraints and the machinery of classical mechanics can be applied to the Polyakov action (in conformal gauge)
\bea
S \sim R^2 \int d^2 \sigma (G_{mn}\partial_a X^m \partial_b X^n \eta^{ab} + {\rm fermions})
\eea
to compute the classical energy and spin of the configuration in the long-string limit (i.e., $\frac{\omega^2-\kappa^2}{\kappa^2} \ll 1$). This leads to the classical relation $E-S=\sqrt{2\lambda} \ln S$, which is insensitive to the fermionic part of the worldsheet action. The quantum corrections on the other hand, depend on the masses of both the bosonic and fermionic fluctuations. To compute them we need to look at the IIA Green-Schwarz action up to quadratic order in the fermions. The details of that are presented in \cite{MR, Fernando, CK}, here we will merely give a line of argument to motivate that a  mismatch between string theory and gauge theory is expected if the result (\ref{bes}) is true.

The first observation is that since the string is restricted to $AdS_3$ just like it was in the $AdS_5\times S^5$ case of Frolov-Tseytlin\cite{Tseytlin}, we can read off the masses of the bosonic fluctuations immediately from \cite{Tseytlin}. There, the masses in the appropriate limit were $m^2=(4\kappa^2,2\kappa^2,2\kappa^2,0,0,0,0,0)$. The last five slots correspond to the massless $S^5$ fluctuations (two of the AdS directions do not show up because of gauge fixing freedom.). Since there is only one transverse direction for the string in $AdS_4$, we can immediately write down the masses in our case to be $m^2=(4\kappa^2,2\kappa^2,0,0,0,0,0,0)$, with the last 6 directions belonging to the $\IC \IP^3$.

Next, we notice that the one loop energy shifty is schematically of the form
\bea
\hspace{-0.5in}\label{int} \int^{\infty}dx \Big[\sum_{{\rm bose}}\sqrt{x^2+m_{{\rm B}}^2} -\sum_{{\rm fermi}}\sqrt{x^2+m_{{\rm F}}^2} \ \Big], \
{\rm which \ is \ finite \ only \ if } \
\sum_{{\rm bosons}}m_{{\rm B}}^2=\sum_{{\rm fermions}}m_{{\rm F}}^2
\eea
By susy, we know there are 8 fermions. Before the ABJM orbifolding, they were in the ${\bf 8}_c$ of $SO(8)$. After the orbifold, they should fall into reps of the $SU(4)\times U(1)$ global symmetry of ABJM. It is a group theoretic fact that ${\bf 8}_c$ decomposes as $  {\bf 6}_{0} \oplus {\bf 1}_{2} \oplus {\bf 1}_{-2}$. Since our classical spinning string does not break the symmetries of $\IC \IP^3$, this means that there will be two groups of massive fermions (one set of six and another of two) each with a distinct mass.  If we parametrize these masses as $m_6^2=\alpha \kappa^2, m_2^2=\beta \kappa^2$, to satisfy (\ref{int}) we will need $\beta=3(1-\alpha)$. Now, by direct computation of the mass shift integral, for the leading term in its $1/\kappa$ expansion to match with the gauge theory result (\ref{bes}), we find that we need $\frac{3}{2}[\alpha \ln \alpha + (1-\alpha) \ln (3-3\alpha) ] =\ln 2$. This is numerically solved by $\alpha= 0.167721...$. But $\alpha$, $\beta$ have to be rational because they arise from rational RR-couplings of the fermions. If one is willing to believe that these rational numbers are reasonably ``small" (i.e., after reducing them to the simplest form, they can be expressed as $\frac{p}{q}$ with $p,q$ both less than, say, 5000) then computer scans can be used to show that there is no solution to $0.167721...=\frac{p}{q}$, anywhere near the precision of the left-hand side.

The last step of the previous reasoning is clearly not rock-solid, but the explicit computation using the Green-Schwarz string can be done, and the result is indeed that there is a mismatch \cite{MR, Fernando, CK}. So the question becomes: what should be modified? There has been a suggestion to modify the regularization scheme on the worldsheet \cite{GM}. Another suggestion was made in \cite{MRT} that the interpolating function that appears in the magnon dispersion relation should have a non-zero one-loop correction. (This correction was assumed to be zero in \cite{GV}.) It would certainly be very interesting to explore these questions further.

\begin{acknowledgement}
  CK thanks the attendees of the 4th RTN meeting at Varna, Bulgaria, and the Superstrings at Cyprus conference, Ayia Napa, Cyprus for questions and comments on talks based on this material. This work is supported in
part by IISN - Belgium (convention 4.4505.86), by the Belgian National
Lottery, by the European Commission FP6 RTN programme MRTN-CT-2004-005104
in which the authors are associated with V. U. Brussel, and by the Belgian
Federal Science Policy Office through the Interuniversity Attraction Pole
P5/27.
\end{acknowledgement}


%

\begin{thebibliography}{[1]}


\bibitem{BFSS}
  T.~Banks, W.~Fischler, S.~H.~Shenker and L.~Susskind,
  Phys.\ Rev.\  D {\bf 55}, 5112 (1997)
  [arXiv:hep-th/9610043].

\bibitem{DB-review}
  D.~S.~Berman,
  Phys.\ Rept.\  {\bf 456}, 89 (2008)
  [arXiv:0710.1707 [hep-th]].

\bibitem{Bagger}
  J.~Bagger and N.~Lambert,
  Phys.\ Rev.\  D {\bf 75}, 045020 (2007)
  [arXiv:hep-th/0611108].
  J.~Bagger and N.~Lambert,
  Phys.\ Rev.\  D {\bf 77}, 065008 (2008)
  [arXiv:0711.0955 [hep-th]].

\bibitem{Gustavsson}
  A.~Gustavsson,
  arXiv:0709.1260 [hep-th].
  A.~Gustavsson,
  JHEP {\bf 0804}, 083 (2008)
  [arXiv:0802.3456 [hep-th]].

\bibitem{Tong}
  N.~Lambert and D.~Tong,
  Phys.\ Rev.\ Lett.\  {\bf 101}, 041602 (2008)
  [arXiv:0804.1114 [hep-th]].
  J.~Distler, S.~Mukhi, C.~Papageorgakis and M.~Van Raamsdonk,
  JHEP {\bf 0805}, 038 (2008)
  [arXiv:0804.1256 [hep-th]].
  S.~Banerjee and A.~Sen,
  arXiv:0805.3930 [hep-th].
  D.~Berenstein and D.~Trancanelli,
  arXiv:0808.2503 [hep-th].

\bibitem{ABJM}
  O.~Aharony, O.~Bergman, D.~L.~Jafferis and J.~Maldacena,
  arXiv:0806.1218 [hep-th].


\bibitem{KM}
  C.~Krishnan and C.~Maccaferri,
  JHEP {\bf 0807}, 005 (2008)
  [arXiv:0805.3125 [hep-th]].

\bibitem{ABC}
  J.~Gomis, G.~Milanesi and J.~G.~Russo,
  JHEP {\bf 0806}, 075 (2008)
  [arXiv:0805.1012 [hep-th]].
  S.~Benvenuti, D.~Rodriguez-Gomez, E.~Tonni and H.~Verlinde,
  arXiv:0805.1087 [hep-th].
  P.~M.~Ho, Y.~Imamura and Y.~Matsuo,
  JHEP {\bf 0807}, 003 (2008)
  [arXiv:0805.1202 [hep-th]].

\bibitem{Mukhi}
  B.~Ezhuthachan, S.~Mukhi and C.~Papageorgakis,
  JHEP {\bf 0807}, 041 (2008)
  [arXiv:0806.1639 [hep-th]].

\bibitem{bl}
  J.~Bagger and N.~Lambert,
  arXiv:0807.0163 [hep-th].

\bibitem{Minahan}
  J.~A.~Minahan and K.~Zarembo,
  JHEP {\bf 0809}, 040 (2008)
  [arXiv:0806.3951 [hep-th]].

\bibitem{stefanski}
B.~j.~Stefanski,
  arXiv:0806.4948 [hep-th].
  G.~Arutyunov and S.~Frolov,
  arXiv:0806.4940 [hep-th].

\bibitem{purespinor}
  G.~Bonelli, P.~A.~Grassi and H.~Safaai,
  JHEP {\bf 0810}, 085 (2008)
  [arXiv:0808.1051 [hep-th]].

\bibitem{BES}
  N.~Beisert, B.~Eden and M.~Staudacher,
  J.\ Stat.\ Mech.\  {\bf 0701}, P021 (2007)
  [arXiv:hep-th/0610251].

\bibitem{GV}
  N.~Gromov and P.~Vieira,
  arXiv:0807.0777 [hep-th].
  C.~Ahn and R.~I.~Nepomechie,
  JHEP {\bf 0809}, 010 (2008)
  [arXiv:0807.1924 [hep-th]].

\bibitem{MR}
  T.~McLoughlin and R.~Roiban,
  arXiv:0807.3965 [hep-th].

\bibitem{Fernando}
  L.~F.~Alday, G.~Arutyunov and D.~Bykov,
  arXiv:0807.4400 [hep-th].

\bibitem{CK}
  C.~Krishnan,
  JHEP {\bf 0809}, 092 (2008)
  [arXiv:0807.4561 [hep-th]].

\bibitem{MRT}
  T.~McLoughlin, R.~Roiban and A.~A.~Tseytlin,
  arXiv:0809.4038 [hep-th].

\bibitem{Shahin}
  M.~M.~Sheikh-Jabbari,
  JHEP {\bf 0409}, 017 (2004)
  [arXiv:hep-th/0406214].
  M.~M.~Sheikh-Jabbari and M.~Torabian,
  JHEP {\bf 0504}, 001 (2005)
  [arXiv:hep-th/0501001].

\bibitem{Basu}
  A.~Basu and J.~A.~Harvey,
  Nucl.\ Phys.\  B {\bf 713}, 136 (2005)
  [arXiv:hep-th/0412310].

\bibitem{Berman}
  D.~S.~Berman and N.~B.~Copland,
  Nucl.\ Phys.\  B {\bf 723}, 117 (2005)
  [arXiv:hep-th/0504044].

\bibitem{Bonelli}
  G.~Bonelli, A.~Tanzini and M.~Zabzine,
  arXiv:0807.5113 [hep-th].
  S.~Terashima,
  JHEP {\bf 0808}, 080 (2008)
  [arXiv:0807.0197 [hep-th]].
  K.~Hanaki and H.~Lin,
  JHEP {\bf 0809}, 067 (2008)
  [arXiv:0807.2074 [hep-th]].
  I.~Jeon, J.~Kim, N.~Kim, S.~W.~Kim and J.~H.~Park,
  JHEP {\bf 0807}, 056 (2008)
  [arXiv:0805.3236 [hep-th]].
  I.~Jeon, J.~Kim, B.~H.~Lee, J.~H.~Park and N.~Kim,
  arXiv:0809.0856 [hep-th].
  J.~KIm and B.~H.~Lee,
  arXiv:0810.3091 [hep-th].
T.~Fujimori, K.~Iwasaki, Y.~Kobayashi and S.~Sasaki,
  arXiv:0809.4778 [hep-th].

\bibitem{non-linear}
  R.~Iengo and J.~G.~Russo,
  arXiv:0808.2473 [hep-th].
  M.~R.~Garousi,
  arXiv:0809.0985 [hep-th].

\bibitem{Bandres}
  M.~A.~Bandres, A.~E.~Lipstein and J.~H.~Schwarz,
  JHEP {\bf 0805}, 025 (2008)
  [arXiv:0803.3242 [hep-th]].
  J.~P.~Gauntlett and J.~B.~Gutowski,
  arXiv:0804.3078 [hep-th].
  G.~Papadopoulos,
  JHEP {\bf 0805}, 054 (2008)
  [arXiv:0804.2662 [hep-th]].

\bibitem{bion}
  N.~R.~Constable, R.~C.~Myers and O.~Tafjord,
  Phys.\ Rev.\  D {\bf 61}, 106009 (2000)
  [arXiv:hep-th/9911136].

\bibitem{const}
  N.~R.~Constable and N.~D.~Lambert,
  Phys.\ Rev.\  D {\bf 66}, 065016 (2002)
  [arXiv:hep-th/0206243].

\bibitem{Raam}
  M.~Van Raamsdonk,
  JHEP {\bf 0805}, 105 (2008)
  [arXiv:0803.3803 [hep-th]].


\bibitem{example}
  B.~E.~W.~Nilsson and C.~N.~Pope,
  Class.\ Quant.\ Grav.\  {\bf 1}, 499 (1984).
  C.~Krishnan and S.~Kuperstein,
  JHEP {\bf 0804}, 009 (2008)
  [arXiv:0801.1053 [hep-th]].

\bibitem{Klebanov}
  S.~S.~Gubser, I.~R.~Klebanov and A.~M.~Polyakov,
  Nucl.\ Phys.\  B {\bf 636}, 99 (2002)
  [arXiv:hep-th/0204051].

\bibitem{Tseytlin}
  S.~Frolov and A.~A.~Tseytlin,
  JHEP {\bf 0206}, 007 (2002)
  [arXiv:hep-th/0204226].

\bibitem{Lee}
  B.~H.~Lee, K.~L.~Panigrahi and C.~Park,
  arXiv:0807.2559 [hep-th].

\bibitem{alday2}
  L.~F.~Alday and J.~M.~Maldacena,
  JHEP {\bf 0711}, 019 (2007)
  [arXiv:0708.0672 [hep-th]].

\bibitem{GM}
N.~Gromov and V.~Mikhaylov,
  arXiv:0807.4897 [hep-th].

\bibitem{Orselli}
  G.~Grignani, T.~Harmark and M.~Orselli,
  Nucl.\ Phys.\  B {\bf 810}, 115 (2009)
  [arXiv:0806.4959 [hep-th]].
  G.~Grignani, T.~Harmark, M.~Orselli and G.~W.~Semenoff,
  JHEP {\bf 0812}, 008 (2008)
  [arXiv:0807.0205 [hep-th]].
  D.~Astolfi, V.~G.~M.~Puletti, G.~Grignani, T.~Harmark and M.~Orselli,
  Nucl.\ Phys.\  B {\bf 810}, 150 (2009)
  [arXiv:0807.1527 [hep-th]].





\end{thebibliography}
%

\end{document}